# Quantitative Transformation for Implementation of Adder Circuits in Physical Systems

Jeff Jones, James G.H. Whiting, Andrew Adamatzky

*Centre for Unconventional Computing*
*University of the West of England*
*Coldharbour Lane*
*Bristol, BS16 1QY, UK.*
*{jeff.jones, james.whiting, andrew.adamatzky}@uwe.ac.uk*

**Abstract**

Computing devices are composed of spatial arrangements of simple fundamental logic gates. These gates may be combined to form more complex adding circuits and, ultimately, complete computer systems. Implementing classical adding circuits using unconventional, or even living substrates such as slime mould *Physarum polycephalum*, is made difficult and often impractical by the challenges of branching fan-out of inputs and regions where circuit lines must cross without interference. In this report we explore whether it is possible to avoid spatial propagation, branching and crossing completely in the design of adding circuits. We analyse the input and output patterns of a single-bit full adder circuit. A simple quantitative transformation of the input patterns which considers the *total number* of bits in the input string allows us to map the respective input combinations to the correct outputs patterns of the full adder circuit, reducing the circuit combinations from a 2:1 mapping to a 1:1 mapping. The mapping of inputs to outputs also shows an incremental linear progression, suggesting its implementation in a range of physical systems. We demonstrate an example implementation, first in simulation, inspired by self-oscillatory dynamics of the acellular slime mould *Physarum polycephalum*. We then assess the potential implementation using plasmodium of slime mould itself. This simple transformation may enrich the potential for using unconventional computing substrates to implement digital circuits.

# 1. Introduction

Classical computing devices are based on spatial arrangements of simple fundamental Boolean logic gates, typically implemented on planar silicon substrates using electronic signalling. The logic gates may be combined to form more complex adding circuits and, ultimately, complete computer systems. Classical computing devices are notable for their miniaturisation and integration, and the speed of operation. Although the dominant form of computation, alternatives to classical computing schemes exist in a number of different physical systems, and indeed proliferate in many living systems. Such non-classical, or unconventional, computing mechanisms take advantage of different computing features within these substrates, such as natural parallelism, morphologically adaptive patterning, self-organisation and emergent behaviour.

Implementing classical Boolean operations in such substrates is not necessarily a good computational 'fit' in that it does not lend itself to the advantages of these substrates (and may indeed exacerbate the disadvantages). Nevertheless, implementation of Boolean operations and more complex circuitry does demonstrate universality that is a characteristic feature of classical schemes. In recent years there have been numerous examples of individual gates and more complex circuits implemented in unconventional substrates, including Belousov-Zhabotinsky medium [14, 8, 38, 7], competing patterns in Game-like cellular automata [23, 22], patterns of crystallisation [1], disordered ensembles of carbon nanotunes [13], liquid crystals [17, 25, 9], organic molecular layers [11], spiking memristors [16], nuclear magentic resonance [12], precipitating reaction-diffusion chemical systems [6], enzymatic systems [37, 21, 26].

Limitations faced by unconventional substrates include the fact that planar arrangements of circuits necessitates crossing or bridging points (where signal 'wires' must cross paths but not interfere with either independent signal). The methods also employ branching points where signals from one path must split into two or more separate paths (or indeed combine two or more signals into fewer paths). The necessity of bridging and branching points generates complex spatial arrangements of circuits and also has implications for the timing of signals within the logic gates. In electronic systems timing may be achieved globally using high-frequency oscillators to provide syn-chronisation but these mechanisms are not suitable in many unconventional substrates. In some substrates, such as circuits based on the migration of the true slime mould *Physarum polycephalum* these timing issues can affect the reliable operation of the gates.

In this article we examine the possibility of removing bridging and branching points from the fundamental operation of a common adding circuit, the one-bit full adder circuit by utilising a transformation of the input bit patterns into a quantitative signal. The quantitative nature of this simple transformation may render it suitable in a wide range of computing substrates. An overview of *Physarum* computing with particular relevance to the approximation of digital logic gates and circuits is described in Section 2. In Section 3 we describe the transformation necessary to implement a quantitative full adder circuit. Section 4 describes experiments using a multi-agent model of slime mould where we assess whether geometric constraints of the organism's environment are sufficient to generate changes in oscillatory frequency required to implement the quantitative adder.

In Section 5 we perform the same experiments using the organism itself to assess the practicality of implementing the quantitative adder using a living substrate. We conclude in Section 6 by stating the contribution of the article, along with the potential advantages and disadvantages of the approach.

## 2. *Physarum*-based Logical Gates and Circuits

The protoplasmic tubes of the acellular slime mould *Physarum polycephalum* have an inherent oscillatory frequency which controls movement by shuttle streaming [15]; this streaming controls the rate and direction of growth for the organism. The mechanism for movement is similar to peristalsis, where contraction causes an advancing frontier of plasmodium; the frequency of peristaltic-like streaming determines direction of movement. *Physarum polycephalum* navigates its environment avoiding negative stimuli such as light while foraging for food and favourable conditions; these external stimuli such as biologically active chemicals [35], light [5] and temperature gradients [15] change streaming frequency of the advancing plasmodium. The streaming contraction causes very low voltage fluctuations, which can be electronically measured [33]. Changes of streaming frequency due to local stimuli have led to the development of biosensors measuring chemicals [35, 31], colours [5] and mechanical stimulation [4].

Foraging behavior has been interpreted as intelligence, thus, presenting the organism with a solvable task, can be interpreted as computation [2]. There has recently been a surge of literature puporting *Physarum* computing, with approaches to plasmodial logic gates based on growth being proposed; Adamatzky has reported protoplasmic tubes act like microfluidic logic gates with mechanical stimulation of tube fragments [10].

Spatial implementation of the full adder circuit using logic gates requires that the path be carefully routed between the individual gates. At certain locations the circuit paths must cross, or bridge, without contaminating other circuit lines. In other instances one circuit must be duplicated to provide input to another gate. Duplication of a circuit is relatively simple in most computing substrates (delay issues notwithstanding) but bridging of circuit paths is not possible, or at least very impractical in certain substrates. This difficulty is particularly manifested in living biocomputing substrates such as the slime mould *Physarum polycephalum*. Individual slime mould logic gates were demonstrated using the growth of *Physarum* across a substrate with chemoattraction [29], ballistic movement and interaction of the slime mould's lamellipodia [3] and photoavoidance [24]. Half adder circuits were designed and tested in simulation in [20] and using the *Physarum* plasmodium [3]; the ballistic slime mould circuits are proved to be substrate independent when implemented in light-sensitive Belousov-Zhabotinsky medium [14]. Difficulties in establishing reliable propagation timing and the unpredictability of behaviour at junctions prevent practical usage of the plasmodium for adding circuits. This is compounded by the slow foraging speed of the plasmodium (measured in tens of hours for a single gate) and suggests that spatial implementation of logical gates and adding circuits may not be the best fit for this unconventional computing substrate.

Alternatives to foraging based processing have recently been suggested. These methods overcomes the limitation of slow computation time due to slow organism propagation speed when using the chemotactic or phototactic logic gates. In previous work we suggested that logic operations may be implemented with single protoplasmic tubes using oscillatory thresholding on this streaming frequency; pre-grown protoplasmic tubes are stimulated with light, chemoattractants and temperature gradients. A threshold is applied to the resultant relative change in streaming frequency which determines the logic gate output; the threshold is determined by the type of gate. The basic Boolean operations *AND*, *OR* and *NOT* were initially reported in 2014 [33] followed by the derived operations *NOR*, *NAND*, *XOR* and *XNOR* [34], indi-cating that the oscillatory threshold logic gates were functionally complete. This improved type of *Physarum* logic gate shortened computation time from a number of hours to a matter of minutes. The article also demonstrated combinational logic circuits by cascading of the individual tube gates; a 2-4 bit decoder, half-adder and full-adder were shown. Despite overcoming the slow computation time of plasmodium propagation based gates, one limitation of the combinational logic of cascaded gates remained; while individual gates produced the correct output up to 92% of the time, cascading gates increased the error as the number of gates used increased [32].

If it were possible to find a way to minimise the cascading of gates in implementations of more complex circuits, this would increase the reliability and practicality of biologically implemented digital circuits. We propose and test such a method in the following sections.

## 3. Transformation of the Single-bit Full Adder Circuit

The full adder circuit enables the addition of two binary inputs, together with a carry input (for example, from a previous calculation). The carry output may be cascaded to other adder circuits so that larger strings of bits may be added. A typical implementation of a full adder circuit is shown in Fig. 1 composed of XOR, AND and OR gates. Note the complex spatial routing of signals from the input channels and from combined outputs of individual logic gates.

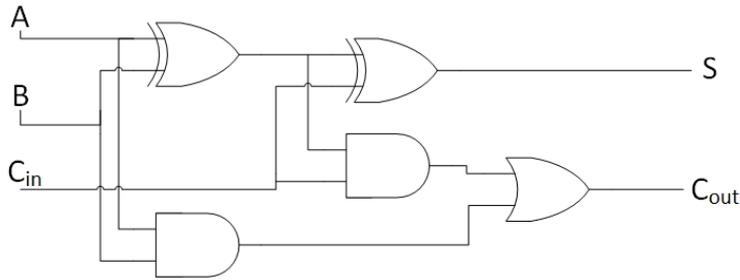

Figure 1: Full adder circuit implemented using spatial arrangements of XOR, AND and OR gates. Note crossing regions (humped lines) and junction regions (joined lines).

*3.1. Quantitative Transformation of Truth Table Input Patterns*

The truth table for the full adder circuit is shown in Table 3.1. In binary terms the full adder may be seen as a 2:1 mapping of the 8 (3-bit) possible input combinations ($X$, $Y$, $C_{in}$) into 4 (2-bit) possible output combinations ($S$, $C_{out}$). If we examine the three inputs to the full adder, together with their corresponding outputs, a pattern can be seen if we ignore the binary values of the inputs, and instead concentrate on the *total* number of '1' bit digits in the input patterns (Table 3.1, column: number of bits). The number of '1' bits in each combination ranges from zero to three, giving four possible combinations, matching the number of output pattern combinations.

Furthermore, combinations of different inputs which share the *same number of bits* all share the same output configuration. For example (referring to Table 3.1) the input patterns which all contain only a single '1' bit (decimal value of patterns 1, 2 and 4) all map to the same output pattern of 0 ($C_{out}$) and 1 ($S$). Similarly, input patterns which contain two '1' bits (decimal value of patterns 3, 5 and 6) all map to the output patterns of 1 ($C_{out}$) and 0 ($S$).

| Decimal | Number of Bits | Input $X$ | Input $Y$ | Input $C_{in}$ | Output $C_{out}$ | Output $S$ |
|---|---|---|---|---|---|---|
| 0 | 0 | 0 | 0 | 0 | 0 | 0 |
| 1 | 1 | 0 | 0 | 1 | 0 | 1 |
| 2 | 1 | 0 | 1 | 0 | 0 | 1 |
| 3 | 2 | 0 | 1 | 1 | 1 | 0 |
| 4 | 1 | 1 | 0 | 0 | 0 | 1 |
| 5 | 2 | 1 | 0 | 1 | 1 | 0 |
| 6 | 2 | 1 | 1 | 0 | 1 | 0 |
| 7 | 3 | 1 | 1 | 1 | 1 | 1 |

Table 1: Single-bit full adder truth table including decimal value of input combinations and the quantitative number of 1 bits in each input combination.

This mapping of quantitatively transformed input pattern bits to output pattern 'bins' is clarified in Table 3.1 which shows the mapping of quantitatively transformed inputs into decimalised interpretations of the output patterns. The result is a 1:1 mapping of (transformed) input to output values. By encoding the number of bits instead of the pattern of bits we have reduced the complexity of the input by half (8 possible combinations to 4).

The mapping removes the need for spatial re-routing of separate components of the adder circuit (for example, in Fig. 1 both inputs 'A' and 'B' are duplicated and passed to more than one gate, whilst input 'C' is fed forward to combine with the output of the first XOR gate and also split to become an input for the second AND gate). Importantly, this mapping of input to output values is also identical in that they both follow an incremental progression. This suggests that it might be possible to use a physical mechanism to perform the computation of the adder circuit. That is, the transformed input could be encoded in a single signal line carrying a weighted non-binary signal.

| Number of Bits | $C_{out}$ | $S$ | Decimalised Outputs |
|:---:|:---:|:---:|:---:|
| 0 | 0 | 0 | 0 |
| 1 | 0 | 1 | 1 |
| 2 | 1 | 0 | 2 |
| 3 | 1 | 1 | 3 |

Table 2: Mapping quantitatively transformed input combinations into truth table outputs. Left column shows the total number of bits in the $X$,$Y$ and $C_{in}$ inputs. Middle columns ($C_{out}$ and $S$) are binary outputs. Right column shows the decimal values of the two binary outputs $C_{out}$ and $S$.

*3.2. Towards Physical Mechanisms for Quantitative Adder Circuits*

A physical implementation of the adder circuit must relate the number of input bits to an increase in some physical value. This value could be changes in light intensity, an increase in pressure, and so on... To return the output to the binary domain the relevant output map 'bin' must be transformed into binary values in order to pass the $S$ and $C_{out}$ on to the next gate (Fig. 2). This conversion adds complexity at the interface of the circuit, but this is balanced by the increased simplicity within the adder circuit itself.

# 4. Implementation of the Mechanism in a Model of *Physarum polycephalum*

As an illustrative example we demonstrate how a multi-agent model of slime mould can be used to implement the quantitative adder. The multi-agent approach is inspired by the simple components of the *Physarum plasmodium* and is composed of a population of simple mobile particles indirectly coupled within a diffusive lattice. Agents loosely correspond to aggregates of overlapping actin filaments. Their collective structure of the population corresponds to the pattern of the tube networks of the plasmodium and the collective movement of particles corresponds to the sol flux within the plasmodium. A full description of the model is given in the appendix and below we concentrate on the transformations of binary values at the inputs to the adder circuit and the subsequent interpretation of the changes in behaviour of the model as the outputs of the circuit.

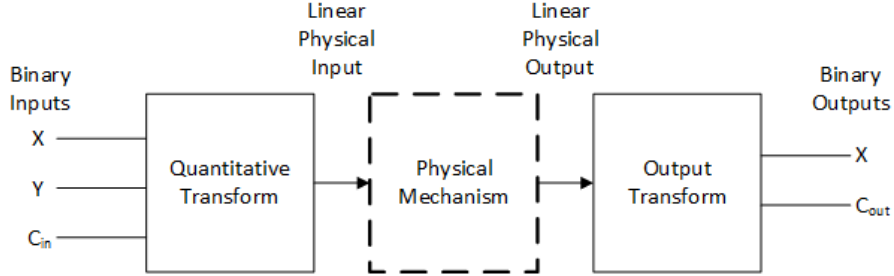

Figure 2: Schematic illustration of quantitative adder mechanism. Binary inputs (left) are transformed into a linear physical quantity which is passed to physical system implementation (dashed) whose output is transformed to a binary representation (right).

We represent the changing physical quantities of the transformed binary inputs by geometrically constraining the model plasmodium within a narrow tube-like arena of $323 \times 20$ pixels (Fig. 3). The model plasmodium (5000 particles) is inoculated within the habitable region of this arena. As the individual particles move they deposit a generic chemoattractant trail within the lattice at their new site. Particles are also attracted by the local concentration of trails. If particles cannot move or collide, no trail is deposited. After a short period of time the initially uniform distribution of trail within the lattice becomes unevenly distributed, causing local oscillations of trail concentration. These oscillatory domains grow and become entrained (see [30] and [27] for a full description of the emergence of oscillatory domains in the model and real plasmodium respectively). A small sampling window $20 \times 20$ pixels at the left of the arena records the mean flux of trails within the lattice within this region at every 5 scheduler steps.

The quantitative input values mapped from 0 to 3 (from inputs of Fig. 3.1) are transformed into a physical representation by only allowing movement of particles which occupy smaller regions of the arena (elongated dashed regions in Fig. 3b, c and d). These regions represent decreasing fractions of the original length of the arena.

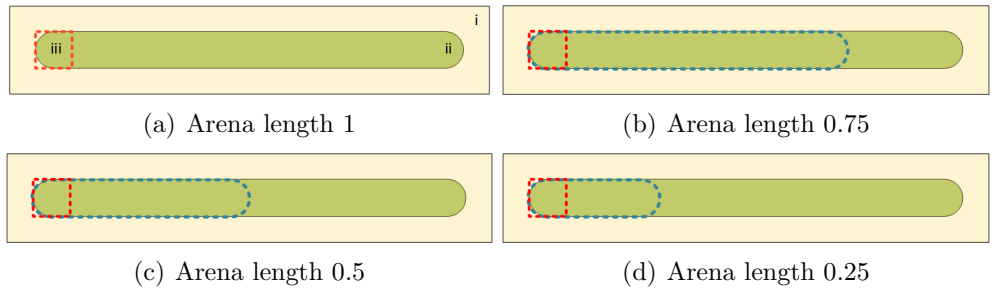

(a) Arena length 1

(b) Arena length 0.75

(c) Arena length 0.5

(d) Arena length 0.25

Figure 3: Geometrically constraining arena of the model plasmodium. a) schematic of model arena with non-habitable area (i), habitable region (ii) and sampling window (iii, dashed red, online) indicated, b,c,d) Constrained habitable region (dashed blue, online) of 0.75, 0.5 and 0.25 the original arena length.

If the geometric constraints of the arena represent the binary input transformation, how is the physical behaviour which generates the individual output bins represented? We can represent this in the model plasmodium by measuring the frequency of oscillations within the arena. The *Physarum* plasmodium, when confined in a suitably shaped arena, exhibits regular oscillations of thickness at each end of the arena [28]. These thickness oscillations emerge from initially random contractions of the plasmodium strand which propel sol throughout the strand. There is a reciprocal relationship between contractile activity and strand thickness (presumably due to the stretch activation phenomenon of the tube strand) as contraction of the tube results in transport of sol away from the contraction site and a subsequent decrease in strand thickness.

The same reciprocal relationship emerges in the model. As particles move they deposit chemoattractant in the lattice (greyscale brightness corresponds to concentration in Fig. 4). The increased flux attracts local particles, causing further increases in population density. Eventually the occupancy in this region is too high to allow free particle movement and flux decreases. At more distant sites the relative decrease in population density (caused by previous particle efflux) allows greater freedom of movement and so flux increases in these regions. The result is a gradual emergence of small oscillatory domains Fig. 4a) which fuse and become entrained Fig. 4b) until a single large oscillation traverses the arena from end to end Fig. 4c, d). The dominant frequency of this stable oscillation Fig. 4e) can be measured by FFT transform and spectral analysis.

When the model plasmodium is geometrically constrained using the patterns in Fig. 3 the dominant oscillatory frequency increases as arena length decreases. An example plot of the oscillatory patterns for decreasing arena length is shown in Fig. 5a. When the oscillation data (10 runs at each arena length fraction) is analysed in the frequency domain, the frequency of oscillations does indeed increase as the arena length decreases. Although the relationship is not perfectly linear (Fig. 5b), a simple thresholding of oscillation frequency should be sufficient to represent the output bins of the quantitative adder.

Although this example uses changing oscillation frequency in a model of *Physarum* plasmodium to represent the physical system, the quantitative concept is applicable to any physical system with approximately linear progression. For example, constraining the geometry of the model plasmodium is suggestive of changing the dominant frequency of a resonant cavity by altering its size.

## 5. Experimental Implementation with Plasmodium of Slime Mould

The plasmodial phase of the organism was cultured on 2% non-nutrient agar, from an initial dried sclerotia, in 9cm diameter Petri-dishes (Sigma-Aldrich, MO, USA). The plasmodium was fed daily with a small number of sterile rolled oat flakes and transplanted to a new Petri-dish every week thus avoiding contamination.

In order to mimic the geometric constraint of the protoplasmic tubes as mentioned in section 4, tubes are grown at fractional lengths between agar hemispheres. Conductive aluminium tape 10mm wide is placed on a 12cm wide square Petri-dish (Sigma-Aldrich, MO, USA) to produce gaps of fixed fractional widths; gap lengths are 3cm, 2.25cm, 1.5cm and 0.75cm which represent fractional length 100%, 75%, 50% and 25% respectively. 1ml 2% agar hemispheres were added to the Petri-dish to facilitate growth (figure 6), and a sterilised oat flake was placed on each hemisphere to maintain the organism as previously desctibed [36]. A sample of plasmodium is placed on one of the pairs of hemispheres and allowed to grow across the gap after which the bio-electric potential was measured. Each tube was measured using a PicoLog ADC-24 high resolution analogue-to-digital data logger (Pico Technology, UK); recording was performed via USB to a laptop installed with the bespoke PicoLog Recorder software. Frequency analysis was performed on the recorded voltage output using custom software written in Matlab R2012A, which consisted of frequency analysis using the Fast-Fourier Transform function and graphical display. For each geometrical length fraction, a total of 10 individual tubes were grown, measured and analysed. The results of frequency analysis for each length is shown in figure 7, oscillation frequency decreases with increasing length; linear regression shows that there is a linear trend through the data with the equation $Y = -0.0015X + 0.0109$ and an $R^2$ value of 0.8107.

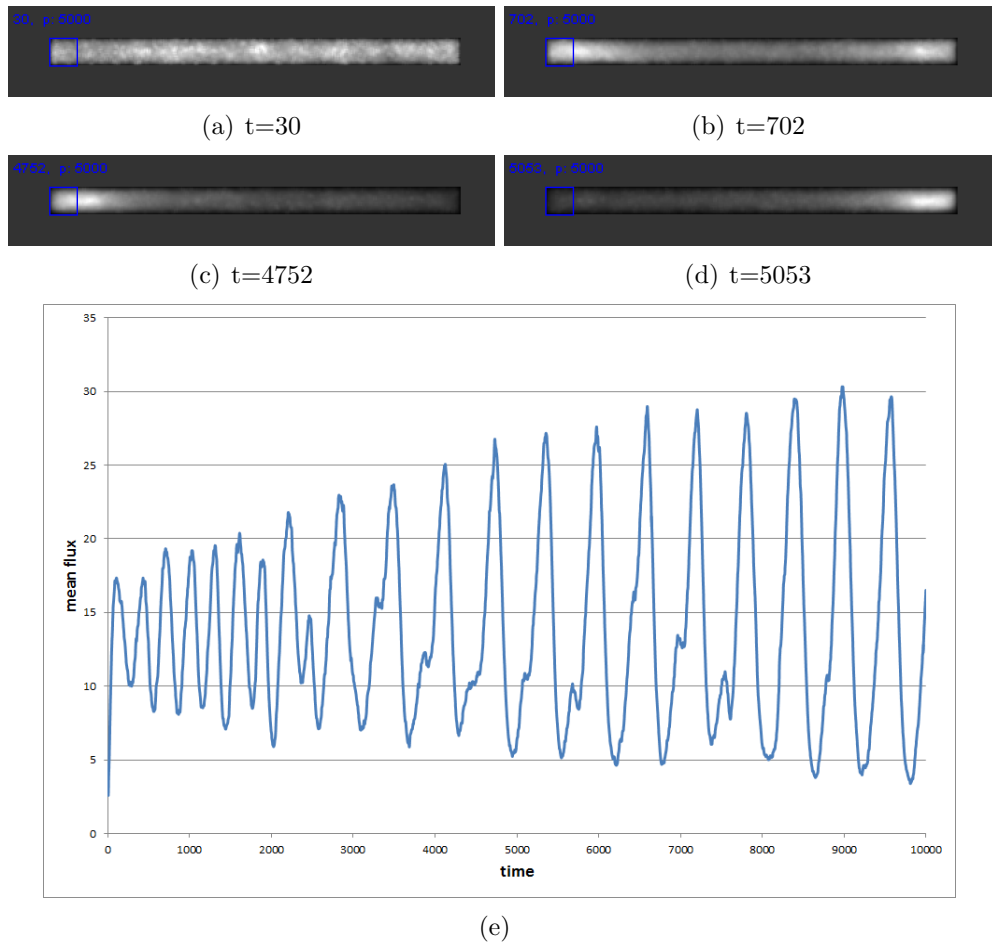

Figure 4: Emergence of regular oscillations of flux within model the plasmodium on full width arena. a) flux of particle trails at t=30 shows relatively uniform distribution (increasing grey brightness corresponds to greater flux), b) at early stages there are two oscillatory domains at each end of the arena, c,d) at later stages there is a single reciprocating oscillatory pattern, e) plot of mean flux within measuring window shows initial oscillatory domains fusing by entrainment at approximately t=2000 to form a regular reciprocating oscillation pattern.

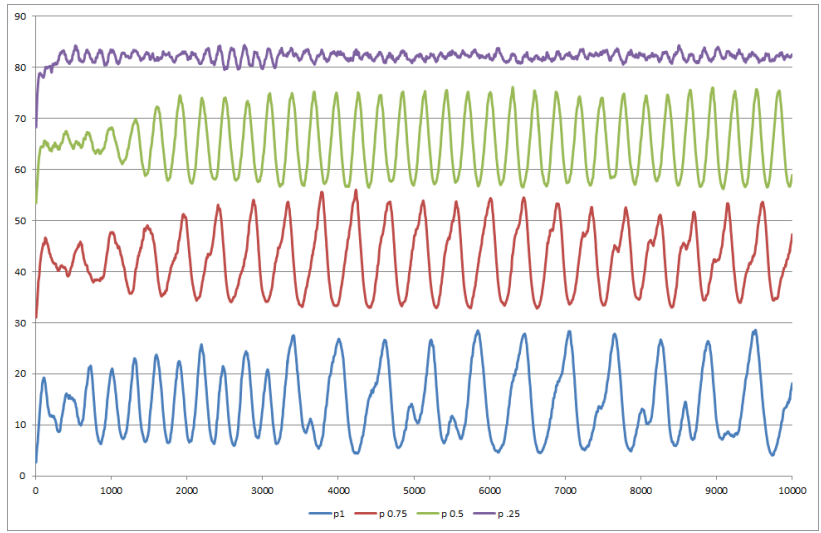

(a)

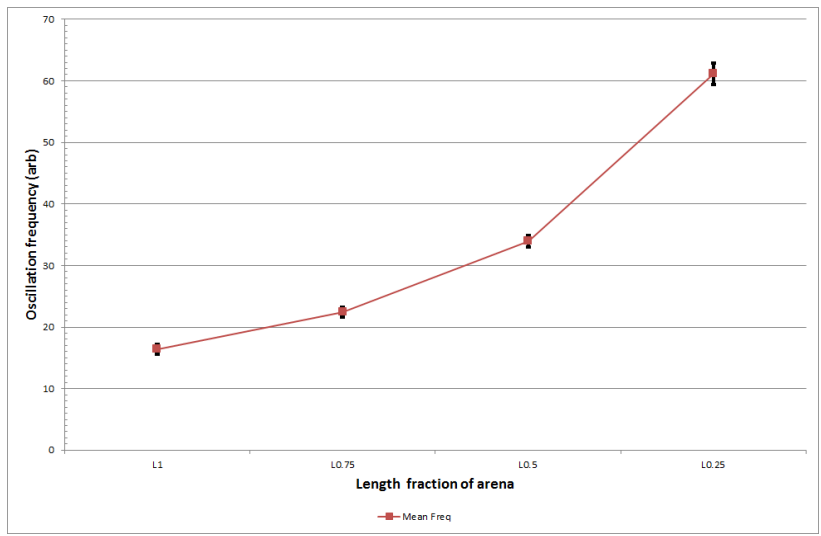

(b)

Figure 5: Oscillation frequency and arena length. a) oscillation frequency increases as arena length decreases, offset plots show example oscillatory activity at arena fractions of (top to bottom) 0.25, 0.5, 0.75 and 1, b) Plot showing increase in oscillation frequency as arena length fraction decreases (10 runs per arena length fraction, standard deviation shown in error bars).

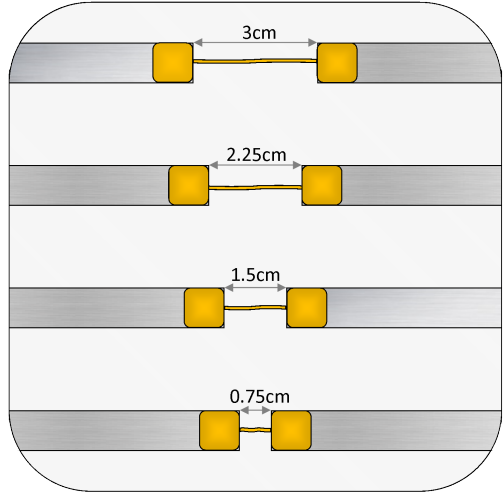

Figure 6: Artistic representation of the Petri-dish used to grow tubes of fractional lengths.

The trend of streaming frequency's dependence on tube length is similar in both the simulations and experimental data; there is a strong negative correlation between length increasing frequency in the experimental data while there is a strong positive correlation between decreasing fraction length and increase in frequency. The experimental results show a linear trend while the model output shows a curved trend; the likely discrepancy is the result of the larger deviation in the experimental data and the ideal length fractions used in the model. The non-linear nature of the model may be observed in different lengths of real protoplasmic tubes should a different set of lengths have been used, however the tubes grow in the experimental conditions were within the range of known protoplasmic tubes which can be reliably and repeatably used. The very similar relationship between the model and experimental data assures validation of the model thus the concept of quantitative transformation for physical systems including but not limited to *Physarum* computation.

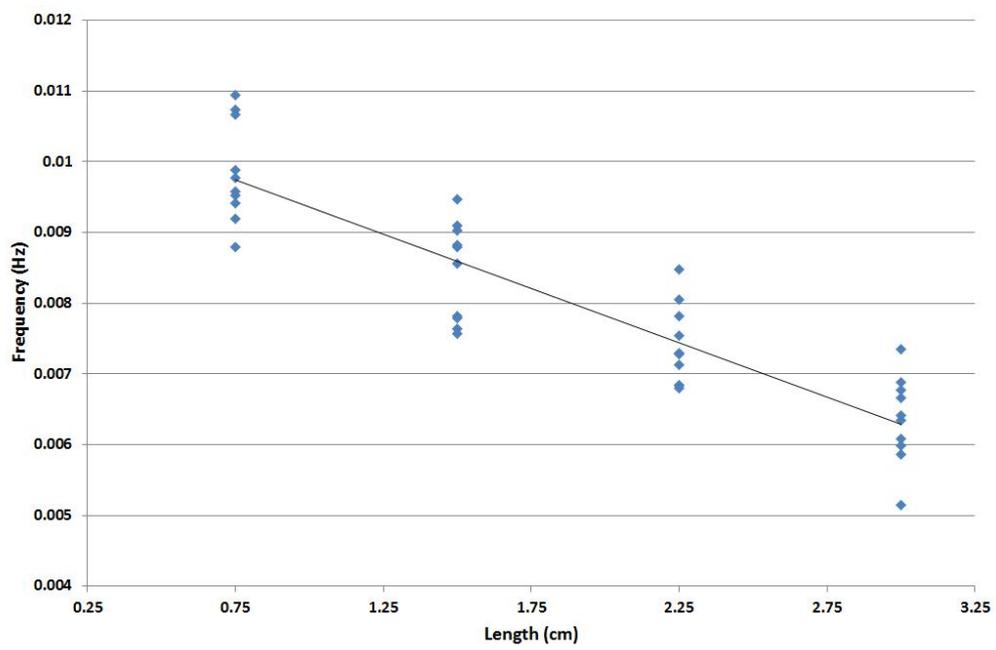

Figure 7: The relationship between shuttle streaming frequency and length of a protoplasmic tube

## 6. Conclusions

We have presented a generalised mechanism for the quantitative computation of a single-bit full adder circuit in order to address the problem of branching points and bridging points which are found in spatially implemented adder circuits. These branch and bridge points render it difficult, or at least impractical, to implement complex circuits in certain unconventional computing substrates, particularly living substrates. The mechanism transforms the input patterns of the adder into physical quantities based upon the number of '1' bits in the input patterns. This transformation reduces the number of input combinations to match the number of output combinations and the resulting mapping also has a linear progression, potentially enabling its implementation in any physical systems whose state changes linearly with input. To implement the adder circuit using a physical system a transformation of inputs to the system, and outputs from the system must be applied. We demonstrate an example of such transformations in both experimental and model representations of slime mould *Physarum polycephalum* in which the input transform is represented by geometrically constraining the (real and virtual) plasmodium. The internal ('physical') mechanism of the adder is performed by the effect of arena length on reciprocating oscillation frequency of the model and the living organism. The output transform is performed by measuring and thresholding the frequency of the oscillations, interpreting each thresholded frequency as the respective output 'bin' from which the Sum and Carry are read. The similarity in model and experimental results validates the concept of the quantitative transformation. The advantages of this quantitative approach to adder design is that it removes the necessity of branching and bridging of signal channels within the circuit. The complex 2D spatial arrangement of the adder circuit is, in this example, transformed to a simple 1D chamber. The simplicity of the mechanism comes, however, with the added complexity of implementing a transform to enable the binary input patterns to be presented to, and read from, the physical system which implements the circuit. The generality of the method, however, suggests it is possible to implement the adder circuit in a number of different physical substrates. In future work it may be possible to apply the approach to more complex circuits and directly pipe the output of one circuit to another in the pipeline without recourse to transforming the values back to the binary domain.

## Acknowledgements

This work was supported by the EU research project "*Physarum* Chip: Growing Computers from Slime Mould" (FP7 ICT Ref 316366)

# 7. Appendix: Particle Model Description

We used a multi-agent approach to generate the *Physarum*-like behaviour. This approach was chosen specifically specifically because we wanted to reproduce the generation of complex behaviour using very simple component parts and interactions, and no special or critical component parts to generate the emergent behaviour. Although other modelling approaches, notably cellular automata, also share these properties, the direct mobile behaviour of the agent particles renders it more suitable to reproduce the flux within the plasmodium. The multi-agent particle model of *Physarum* used to generate the oscillatory dynamics [19] uses a population of coupled mobile particles with very simple behaviours, residing within a 2D diffusive lattice. The lattice stores particle positions and the concentration of a local diffusive factor referred to generically as chemoattractant. Particles deposit this chemoattractant factor when they move and also sense the local concentration of the chemoattractant during the sensory stage of the particle algorithm. Collective particle positions represent the global pattern of the material. The model runs within a multi-agent framework running on a Windows PC system. Performance is thus influenced by the speed of the PC running the framework. The particles act independently and iteration of the particle population is performed randomly to avoid any artifacts from sequential ordering.

*7.1. Generation of Virtual Plasmodium Cohesion and Oscillatory Rhythm*

The behaviour of the particles occurs in two distinct stages, the sensory stage and the motor stage. In the sensory stage, the particles sample their local environment using three forward biased sensors whose angle from the forwards position (the sensor angle parameter, SA), and distance (sensor offset, SO) may be parametrically adjusted (Fig. 8a). The offset sensors generate local indirect coupling of sensory inputs and movement to generate the cohesion of the material. The SO distance is measured in pixels and in this article we used an SO value of 15. During the sensory stage each particle changes its orientation to rotate (via the parameter rotation angle, RA) towards the strongest local source of chemoattractant (Fig. 8b). Variations in both SA and RA parameters have been shown to generate a wide range of reaction-diffusion patterns [18] and for these experiments we used SA 90° and RA 22.5° which results in strong oscillatory domains emerging within the population. After the sensory stage, each particle executes the motor stage and attempts to move forwards in its current orientation (an angle from 0–360°) by a single pixel forwards. Each lattice site may only store a single particle and particles deposit chemoattractant into the lattice (5 units per step) only in the event of a successful forwards movement. If the next chosen site is already occupied by another particle move is abandoned and the particle does not deposit any attractant. To generate emergent oscillatory dynamics a new direction is not selected if the move is not successful (see [30] for more information about the emergence of oscillatory dynamics within the model).

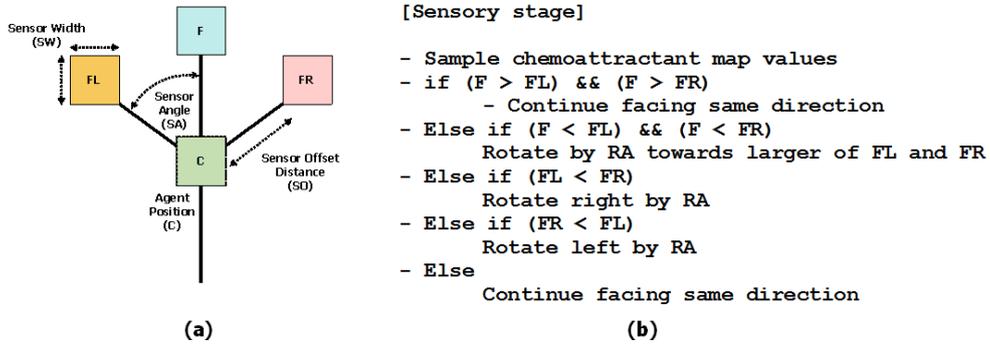

Figure 8: Architecture of a single particle of the virtual material and its sensory algorithm. (a) Morphology showing agent position 'C' and offset sensor positions (FL, F, FR), (b) Algorithm for particle sensory stage.

7.2. Problem Representation

The population resides within a 2D lattice. The lattice comprises a horizontal arena of $360 \times 66$ pixels, within which is a habitable rectangular region sized $300 \times 20$ pixels. Particle movement is bounded by the border of this region. Diffusion of the chemoattractant trails deposited by the particles is implemented by a simple mean filter kernel of size $3 \times 3$ which is damped to reduce diffusion distance by multiplying the mean value of each cell by 0.99. Chemoattractant is deleted when it diffuses outside habitable regions. Geometric constraint of arena size is performed by reducing the habitable size of the arena by fractions of 1, 0.75, 5 and 0.25. Any particle present in regions which are no longer habitable are 'frozen' in space, by which we mean that their motor and sensory behaviours are not implemented. The flux of attractant is measured as the mean value of chemoattractant trail within a $20 \times 20$ window at the left side of the arena. This window is sampled every 5 scheduler steps and the results saved for analysis. Frequency analysis was performed using Sigview spectrum analyser (SignalLAb software).